\documentclass[aps,PRB,reprint,superscriptaddress,preprintnumbers]{revtex4-2}
\usepackage[T1]{fontenc}
\usepackage{times}
\usepackage{graphicx,color}
\usepackage{amsfonts,amsmath,amssymb,amsbsy}
\usepackage[colorlinks=true, linkcolor=blue, citecolor=blue, urlcolor=blue]{hyperref}
\usepackage{float}

\begin{document}

\title{Supersymmetric non-Hermitian topological interface laser}
\author{Motohiko Ezawa}
\affiliation{Department of Applied Physics, The University of Tokyo, 7-3-1 Hongo, Tokyo
113-8656, Japan}
\author{Natsuko Ishida}
\affiliation{Research Center for Advanced Science and Technology, The University of
Tokyo, 4-6-1 Komaba, Tokyo 113-8656, Japan}
\author{Yasutomo Ota}
\affiliation{Research Center for Department of Applied Physics and Physico-Informatics,
Keio University, 3-14-1 Hiyoshi, Japan}
\author{Satoshi Iwamoto}
\affiliation{Research Center for Advanced Science and Technology, The University of
Tokyo, 4-6-1 Komaba, Tokyo 113-8656, Japan}

\begin{abstract}
We investigate laser emission at the interface of a topological and trivial
phases with loss and gain. The system is described by a Su-Schrieffer-Heeger
model with site-dependent hopping parameters. We study numerically and
analytically the interface states. The ground state is described by the
Jackiw-Rebbi mode with a pure imaginary energy, reflecting the
non-Hermiticity of the system. It is strictly localized only at the A sites.
We also find a series of analytic solutions of excited states based on SUSY
quantum mechanics, where the A and B sites of the bipartite lattice form
SUSY partners. We then study the system containing loss and gain with
saturation. The Jackiw-Rebbi mode is extended to a nonlinear theory, where B
sites are also excited. The relative phases between A and B sites are fixed,
and hence it will serve as a large area coherent laser.
\end{abstract}

\date{\today }
\maketitle

\section{Introduction}

Topological physics is one of the most exciting fields\cite{Hasan,Qi}. The
Su-Schrieffer-Heeger (SSH) model is a simplest example of topological
insulators\cite{SSH}. The topological phase is characterized by the
emergence of zero-energy states at the edges of a sample. A zero-energy
state emerges also at an interface between a topological phase and a trivial
phase, which is called a topological interface state. The Jackiw-Rebbi
solution\cite{JR} is an analytic solution for the topological interface
state. Now, non-Hermitian topological physics is an emerging field. The
Jackiw-Rebbi solution seems to be not valid because the energy of the
topological interface state is nonzero in general.

Topological photonics is an ideal playground of studying topological physics%
\cite%
{KhaniPhoto,Hafe2,Hafezi,WuHu,TopoPhoto,Ozawa16,Ley,KhaniSh,Zhou,Ozawa,Ota19,OzawaR,Hassan,Ota,Li,Yoshimi,Kim,Iwamoto21}%
. Topological laser is one of the most successful applications of
topological physics\cite%
{Harrari,Bandres,Schome,Weimann,Jean,Ota18,Parto,Zhao,Malzard,MalzardOpt,Zhong}%
. A strong lasing from a single coherent mode is possible due to a
topological edge or interface state. In topological photonics, loss is
inevitable and hence leading to non-Hermitian topological physics\cite%
{LFeng,GanaRev}. We need to add a gain in order to obtain a laser.
Especially, a topological interface laser has enabled a large area coherent
lasing by using a smooth interface\cite{Ishida}.

In this paper, in order to understand laser emission at the interface
between a topological and trivial phases, we analyze a non-Hermitian SSH
model first by including linear loss and gain terms. We solve numerically a
set of nonlinear differential equations. We also make an analytical study of
the Jackiw-Rebbi mode to describe the topological interface state, upon
which we construct a series of excitation states at the interface based on
supersymmetric (SUSY) quantum mechanics generalized to a non-Hermitian
system. We call them SUSY Jackiw-Rebbi modes because they preserve SUSY
although the original Jackiw-Rebbi mode breaks SUSY. Not only the
topological interface state but also the SUSY Jackiw-Rebbi modes are shown
to have pure imaginary energies. Here, SUSY partners are formed by the A and
B sites of the bipartite lattice, where only A sites are excited in the
original Jackiw-Rebbi mode. We confirm that the analytical solutions well
coincide with numerical solutions. Next, we include a saturation term to the
gain, which is a nonlinear term. Such a system well describes a large area
stable laser emission from an interface of a topological system. The
Jackiw-Rebbi topological mode is solely stimulated in laser emission. We
extend the Jackiw-Rebbi mode to the nonlinear regime. Excitations at B sites
are induced in the Jackiw-Rebbi mode by a nonlinear effect, where the
wavefunction at B sites is fixed to be pure imaginary. The relative phases
between the saturated wavefunctions at the A and B sites are fixed. Since
the Jackiw-Rebbi mode extends over a wide region around the interface, it
will give a large area coherent laser.

\section{Model}

We investigate the dynamics of a laser system governed by\cite{Harrari} 
\begin{equation}
i\frac{d\psi _{n}}{dt}=\sum_{nm}M_{nm}\psi _{m}-i\gamma \left( 1-\chi \frac{%
\left( 1-\left( -1\right) ^{n}\right) /2}{1+\left\vert \psi _{n}\right\vert
^{2}/\eta }\right) \psi _{n},  \label{Master}
\end{equation}%
with a site dependent hopping matrix%
\begin{align}
M_{nm}=& \kappa _{A,n}\left( \delta _{2n,2m-1}+\delta _{2m,2n-1}\right) 
\notag \\
& +\kappa _{B}\left( \delta _{2n,2m+1}+\delta _{2m,2n+1}\right) ,
\label{HoppiMatrix}
\end{align}%
where $\psi _{n}$ is the amplitudes at the site $n$, where $n=1,2,3,\cdots
,N $ in the system composed of $N$ sites; $\gamma $ represents the loss in
each resonator; $\gamma \chi $ represents the amplitude of the optical gain
via stimulated emission induced only at the odd site; $\eta $ represents the
nonlinear saturation constant\cite{Harrari}. All these parameters are taken
positive semidefinite. The lattice structure of the SSH model is bipartite,
where the odd and even sites are called the A and B sites, respectively. The
system turns out to be a linear model in the limit $\eta \rightarrow \infty
. $ On the other hand, $\gamma $\ controls the non-Hermicity, where the
system is Hermitian for $\gamma =0$.

\begin{figure}[t]
\centerline{\includegraphics[width=0.48\textwidth]{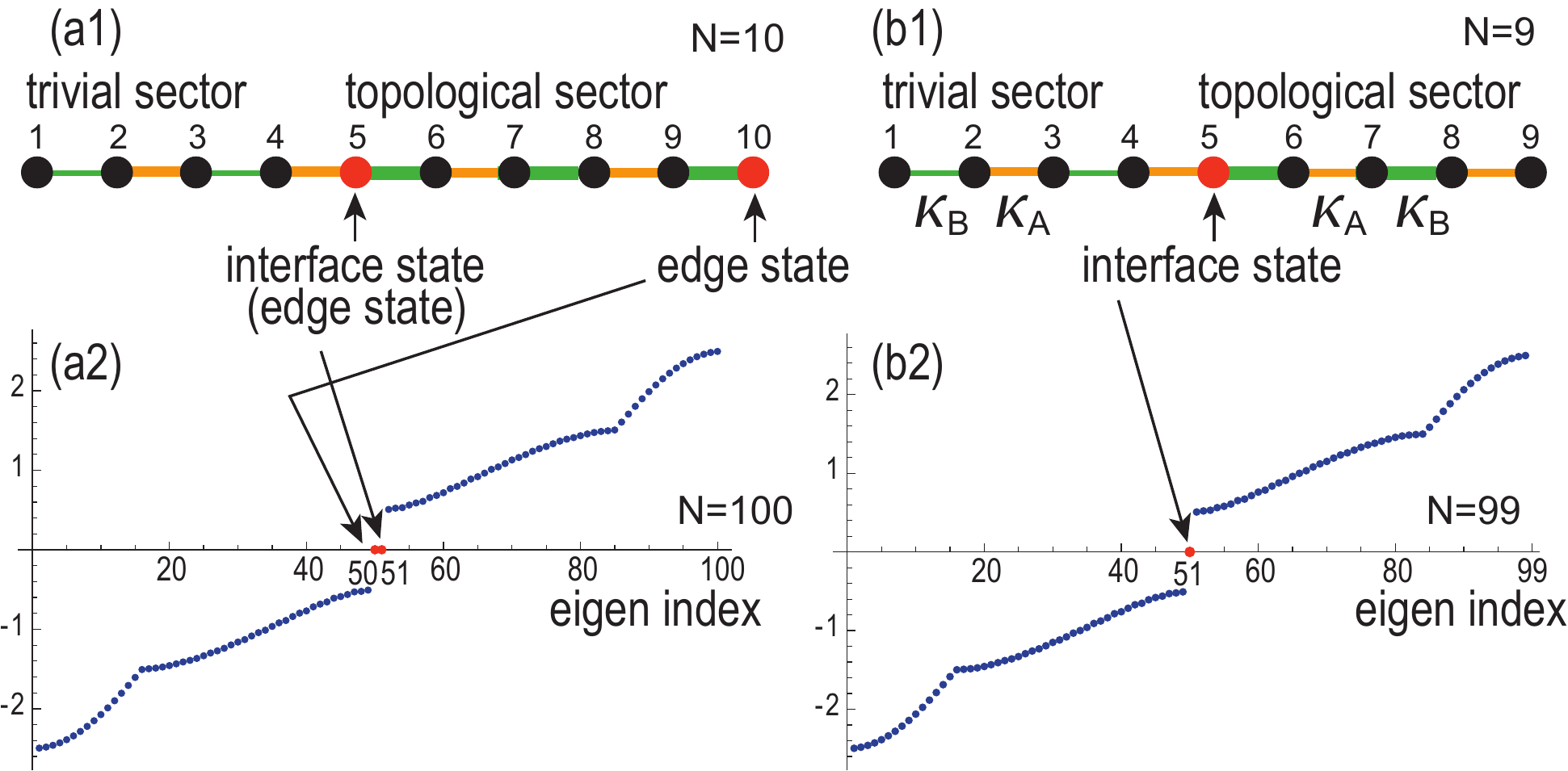}}
\caption{(a1), (b1) Illustration of the interface (marked in red) in the SSH
chain for $N=10$ and $9$. (a1) Topological edge states (marked in red)
appear at the two edges of a topological sector. (b1) The topological edge
state is absent at the edge of a sample when $N$ is odd. The topological
state emerges only at the interface. (a2), (b2) Energy spectrum (vertical
axis) of the SSH model as a function of the eigen index (horizontal axis)
for $N=100$ and $99$, where the eigen index is sorted in the increasing
order of the energy. Two and one zero-energy topological states (marked in
red) emerge in the SSH chain with $N=10$ and $9$. The structure of kinks at $%
p=16$ and $p=84$ is due to the difference of the band width between the
topological and the trivial sectors. We have set $\protect\lambda =0.5$.}
\label{FigEdgeInterface}
\end{figure}

The hopping amplitudes are explicitly given by%
\begin{equation}
\kappa _{A,n}=\kappa \left( 1+\lambda \tanh \frac{n-n_{\text{IF}}+1/2}{\xi }%
\right) ,\quad \kappa _{B}=\kappa ,  \label{EqA}
\end{equation}%
with $\lambda >0$, where $n_{\text{IF}}$ is the smallest odd number larger
than or equal to $N/2$. Then, $n-n_{\text{IF}}+1/2>0$ for $n\geq n_{\text{IF}%
}$, and $n-n_{\text{IF}}+1/2<0$ for $n<n_{\text{IF}}$. We call the site $%
n=n_{\text{IF}}$ the interface of the chain. See Fig.\ref{FigEdgeInterface}%
(a1) and (b1) for an illustration in the case of $N=10$ and $9$.

The explicit equations for a finite chain with length $N$\ follow from Eq.(%
\ref{Master}) as%
\begin{align}
i\frac{d\psi _{2n-1}}{dt}=& \kappa _{B}\psi _{2n-2}+\kappa _{A,n}\psi _{2n} 
\notag \\
& -i\gamma \left( 1-\frac{\chi }{1+\left\vert \psi _{2n-1}\right\vert
^{2}/\eta }\right) \psi _{2n-1},  \label{SSH1} \\
i\frac{d\psi _{2n}}{dt}=& \kappa _{B}\psi _{2n+1}+\kappa _{A,n}\psi
_{2n-1}-i\gamma \psi _{2n}.  \label{SSH2}
\end{align}%
We solve this set of equations together with the initial condition%
\begin{equation}
\psi _{n}\left( t=0\right) =\delta _{n,n_{\text{IF}}}.  \label{IniCon}
\end{equation}%
This is a quench dynamics starting from the interface site by giving an
input to it initially. The initial input triggers the gain effect in Eq.(\ref%
{SSH1}) because $n_{\text{IF}}$ is an odd number.

\section{Linear theory}

We start with the linear model ($\eta \rightarrow \infty $). Then, Eq.(\ref%
{Master}) is reduced to%
\begin{equation}
i\frac{d\psi _{n}}{dt}=\sum_{m}\widetilde{M}_{nm}\psi _{m},  \label{TopoDyna}
\end{equation}%
where%
\begin{equation}
\widetilde{M}_{nm}=\overline{M}_{nm}-i\gamma \left( 1-\frac{\chi }{2}\right)
\delta _{nm},  \label{HoppiMatrixA}
\end{equation}%
with%
\begin{equation}
\overline{M}_{nm}=M_{nm}-i\gamma \chi \frac{\left( -1\right) ^{n}}{2}\delta
_{nm}.  \label{HoppiMatrixB}
\end{equation}%
Since $\widetilde{M}_{nm}$ and $\overline{M}_{nm}$ are different only by a
c-number term, they describe the identical physics. Hereafter, we use $%
\widetilde{M}_{nm}$ for the study of dynamics and $\overline{M}_{nm}$ for
the analytical study of the system.

\subsection{Topological edge and interface states}

\subsubsection{SSH model}

We analyze the SSH model $M_{nm}$ by taking the negligible penetration depth
($\xi \rightarrow 0$). Then, Eq.(\ref{EqA}) amounts to

\begin{align}
\kappa _{A,n} &=\kappa \left( 1+\lambda \right) \quad \text{for}\quad n\geq
n_{\text{IF}},  \notag \\
\kappa _{A,n} &=\kappa \left( 1-\lambda \right) \quad \text{for}\quad n<n_{%
\text{IF}}.  \label{EqB}
\end{align}%
The hopping amplitudes are constant $\kappa _{A,n}=\kappa \left( 1+\lambda
\right) $\ for the segments with $n\geq n_{\text{IF}}$,\ while they are
constant $\kappa _{A,n}=\kappa \left( 1-\lambda \right) $\ for the segments
with $n<n_{\text{IF}}$, separately. Note that $\kappa _{B}=\kappa $. The
hopping matrix $M_{nm}$ defines the SSH model in each segment.

The SSH model with constant hopping amplitudes $\kappa _{A}$ and $\kappa
_{B} $ has a topological phase for $\kappa _{A}<\kappa _{B}$ and the trivial
phase for $\kappa _{A}>\kappa _{B}$. The topological phase is characterized
by the emergence of zero-energy states at the edges of a finite chain, as
demonstrated numerically in Fig.\ref{FigEdgeInterface}(a2) for $N=100$. This
is the standard bulk-edge correspondence. It is illustrated in Fig.\ref%
{FigEdgeInterface}(a1) for $N=10$. See Appendix for details.

There is an intriguing phenomenon in the SSH model with respect to the
even-odd effect of the number of the sites within the chain\cite{Zhao,Ishida}%
. We may remove the edge site at $n=N$ from an SSH chain with even $N$ to
obtain an SSH chain with odd total number $N-1$. See an illustration in Fig.%
\ref{FigEdgeInterface}(a1) and (b1), where two chains with $N=10$ and $9$
are shown. We demonstrate numerically that there is only one zero-mode state
in the odd chain with $N=99$ in Fig.\ref{FigEdgeInterface}(b2), which is the
topological interface state illustrated in Fig.\ref{FigEdgeInterface}(a2).
This is also a bulk-edge correspondence. Recall that the topological number
is defined for the unit cell of the bulk.

In the rest of this work, we focus on the topological interface state by
taking an SSH chain with odd $N$. Furthermore, we do not take the limit $\xi
\rightarrow 0$ any longer.

\begin{figure}[t]
\centerline{\includegraphics[width=0.48\textwidth]{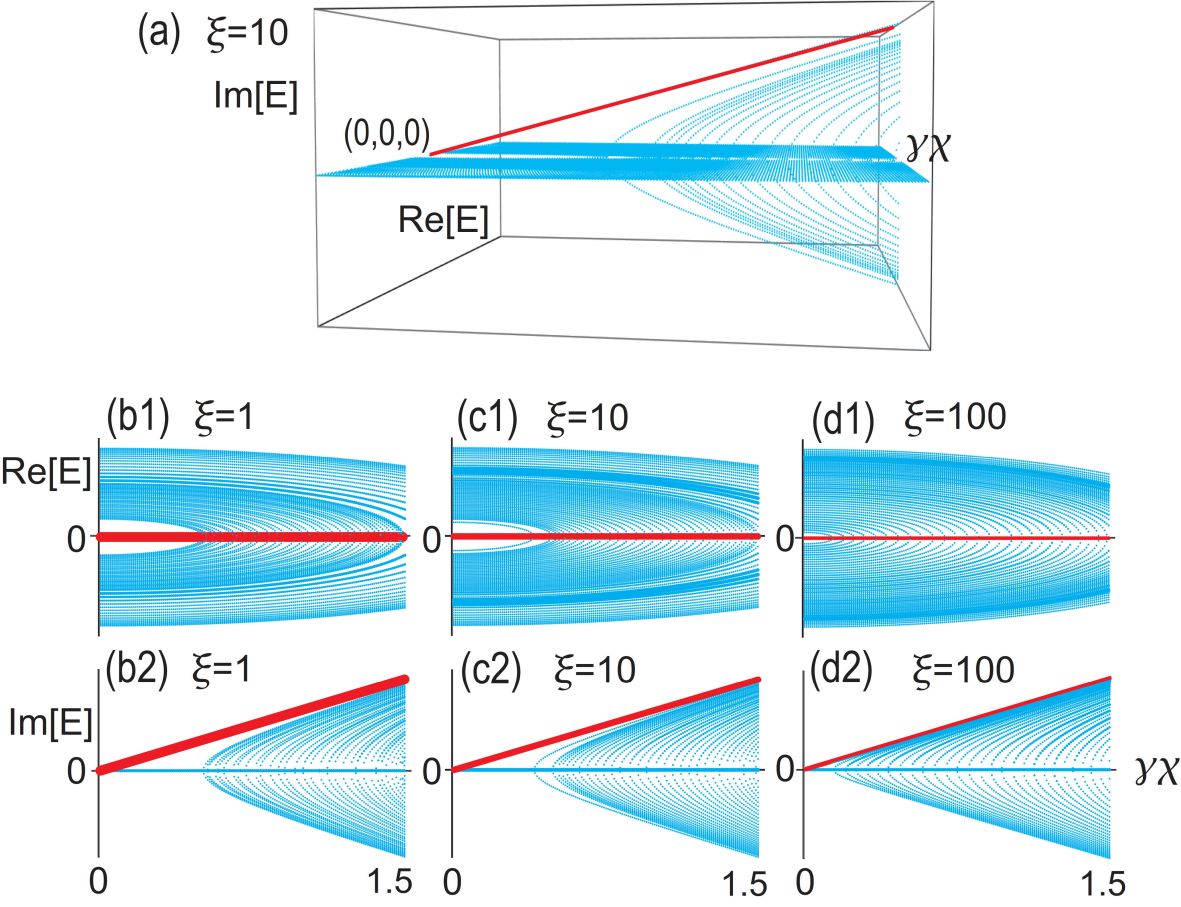}}
\caption{(a) Energy spectrum in the $(\protect\gamma\protect\chi ,$Re$[%
\overline{E}],$Im$[\overline{E}])$ space for $\protect\xi =10$, where $%
\protect\gamma\protect\chi $ stands for the gain $(0< \protect\gamma\protect%
\chi <1.5)$. (b1), (c1), (d1) Energy spectrum in the $(\protect\gamma\protect%
\chi ,$Re$[\overline{E}])$ plane for $\protect\xi =1,10,100$. (b2), (c2),
(d2) Energy spectrum in the $(\protect\chi ,$Im$[\overline{E}])$ plane for $%
\protect\xi =1,10,100$. The red line represents the topological interface
state, whose energy is pure imaginary. The width of the line is proportional
to the local density of states. The interface state is well separated from
(almost touched to) the bulk spectrum for $\protect\xi =1,10$ ($\protect\xi %
=100$). We have set $\protect\gamma =0.1$ and $\protect\lambda =0.5$. We
have used the chain with $N=99$. }
\label{FigEneIm}
\end{figure}

\subsubsection{Non-Hermitian SSH model}

We investigate the system $\overline{M}_{nm}$ with a finite loss ($\gamma
\neq 0$) and gain ($\gamma \chi \neq 0$). Diagonalizing the hopping matrix $%
\overline{M}_{nm}$ in Eq.(\ref{HoppiMatrixA}) numerically, we obtain the
energy spectrum $\overline{E}$ as a function of $\chi $\ while setting $%
\gamma =0.1$. We show the results in the $(\chi ,\text{Re}[\overline{E}],%
\text{Im}[\overline{E}])$ space for $\xi =10$ in Fig.\ref{FigEneIm}(a). See
also Fig.\ref{FigEneIm}(c1) and (c2) for its cross section at Im$[\overline{E%
}]=0$ and Re$[\overline{E}]=0$, respectively. We clearly observe a straight
line passing through the point $(0,0,0)$ in the $(\chi ,\text{Re}[\overline{E%
}],\text{Im}[\overline{E}])$ space, which represents the energy of the
topological interface state we have just discussed.

Similarly, we show the energy spectrum for $\xi =1$ and $100$ in Fig.\ref%
{FigEneIm}(b1), (b2), (d1) and (d2). We also find a straight line passing
through the point $(0,0,0)$ in the $(\chi ,\text{Re}[\overline{E}],\text{Im}[%
\overline{E}])$ space.

The energy of the topological interface state is well fitted for any system
parameters by the formula%
\begin{equation}
\overline{E}_{\text{IF}}=i\bar{\gamma}\quad \text{with}\quad \bar{\gamma}%
=\gamma \chi /2.  \label{EneIF}
\end{equation}%
The eigenvalue (\ref{EneIF}) and the associated eigenfunction are derived as
a Jackiw-Rebbi solution later in Section \ref{SecDirac}: See Eq.(\ref{JRa}).

In addition, we observe a band-edge mode\cite{Ishida} between the interface
mode and the bulk spectrum for $\xi =10$. In the case of $\xi =100$, in
addition to the band-edge mode, there are many modes with almost equal
spacing and characterized by their pure imaginary energies. We call them
SUSY Jackiw-Rebbi modes, with respect to which we discuss based on the SUSY
quantum mechanics in Section \ref{SecDirac}.

\begin{figure}[t]
\centerline{\includegraphics[width=0.48\textwidth]{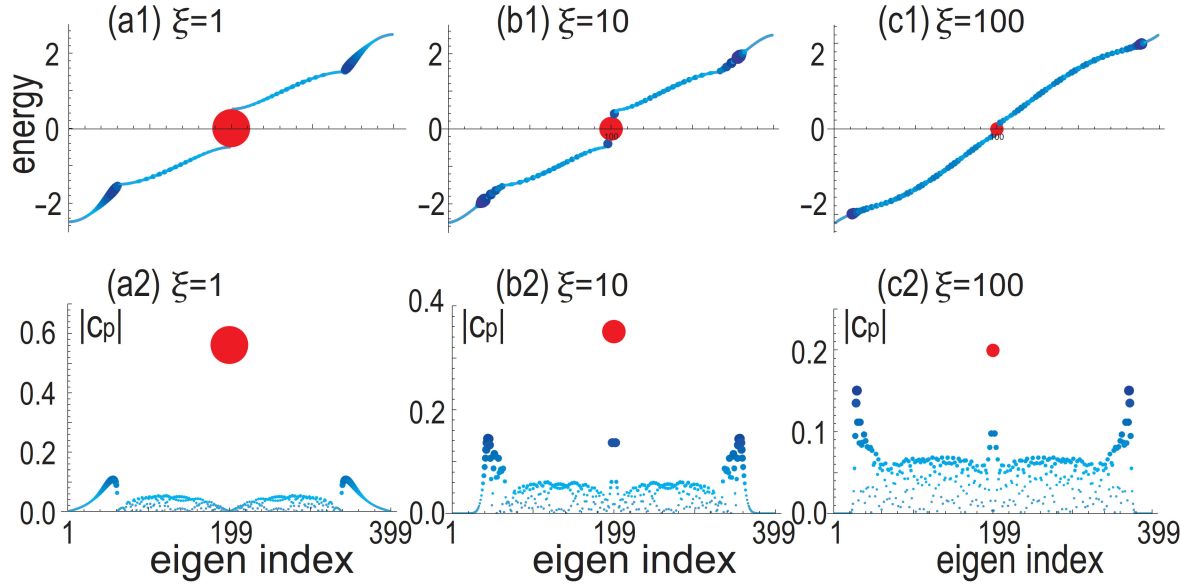}}
\caption{(a1), (b1), (c1) Energy $E_{p}$ and (a2), (b2), (c2) the component $%
|c_{p}|$ as functions of the eigen index $p$. A red large disk indicates the
topological interface state. On the other hand, cyan small disks indicate
the bulk states. The size of a disk is proportional to the local density of
states. It becomes smaller for larger $\protect\xi $ because the interface
mode becomes broader. The horizontal axis is the eigen index. (a1), (a2) $%
\protect\xi =1$; (b1), (b2) $\protect\xi =10$; (c1), (c2) $\protect\xi =100$%
. We have set $N=399$, $n_{\text{IF}}=199$ and $\protect\gamma =0$.}
\label{FigIniExpand}
\end{figure}

\subsection{Dynamics}

The quench dynamics is a powerful tool to distinguish topological phase even
for nonlinear systems\cite{TopoToda,MechaRot,NLPhoto,TopoLaser}. Before
analyzing the dynamics of the system, it is convenient to study the
eigenvalues and the eigenfunctions of the hopping matrix $\widetilde{M}_{nm}$
given by Eq.(\ref{HoppiMatrixA}). We diagonalize it as 
\begin{equation}
\widetilde{M}\phi _{p}=\widetilde{E}_{p}\phi _{p},  \label{EigenA}
\end{equation}%
where $p$ labels the eigen index, $1\leq p\leq N$, and $\phi _{p}$ is the
eigenfunction. We show the eigenvalues $\widetilde{E}_{p}$ in Fig.\ref%
{FigIniExpand}(a1), (b1) and (c1). Let the wavefunction of the topological
interface state be $\phi _{\text{IF}}$. Its eigenvalue is%
\begin{equation}
\widetilde{E}_{\text{IF}}=\overline{E}_{\text{IF}}-i\gamma \left( 1-\frac{%
\chi }{2}\right) \delta _{nm}=i\gamma \left( \chi -1\right) ,  \label{EneIFB}
\end{equation}%
with the use of Eq.(\ref{HoppiMatrixA}) and Eq.(\ref{EneIF}).

Decoupled equations follow from Eq.(\ref{TopoDyna}) for the eigenfunctions,%
\begin{equation}
i\frac{d\phi _{p}}{dt}=\widetilde{E}_{p}\phi _{p},
\end{equation}%
whose solutions are given by%
\begin{equation}
\phi _{p}\left( t\right) =\exp \left[ -it\widetilde{E}_{p}\right] \phi _{p}.
\label{LSol}
\end{equation}%
In particular, for the topological interface state, we have%
\begin{equation}
\phi _{\text{IF}}\left( t\right) =\exp \left[ \gamma \left( \chi -1\right) t%
\right] \phi _{\text{IF}},  \label{TimeEvoIF}
\end{equation}%
with the use of Eq.(\ref{EneIFB}). It has no dynamics for $\gamma =0$ or $%
\chi =1$. On the other hand, it grows exponentially for $\chi >1$.

The initial state (\ref{IniCon}) is expanded in terms of the eigenfunctions
as 
\begin{equation}
\psi _{n}\left( t=0\right) =\delta _{n,n_{\text{IF}}}=\sum_{p}c_{p}\phi _{p}.
\label{Expand}
\end{equation}%
We show the coefficient $|c_{p}|$ in Fig.\ref{FigIniExpand}(a2), (b2) and
(c2), which is determined by%
\begin{equation}
c_{p}=\sum_{n}\delta _{n,n_{\text{IF}}}\phi _{p}.
\end{equation}%
It is the overlap between the initial state (\ref{IniCon}) and the
eigenstate $\phi _{p}$. Such an overlap for the topological interface $\phi
_{\text{IF}}$ is $|c_{\text{IF}}|$, which is found large for $\xi =1$ but
small for $\xi =100$ in Fig.\ref{FigIniExpand}. This is because the
topological interface state is strictly localized at the interface for small 
$\xi $, but broad for large $\xi $.

\begin{figure}[t]
\centerline{\includegraphics[width=0.48\textwidth]{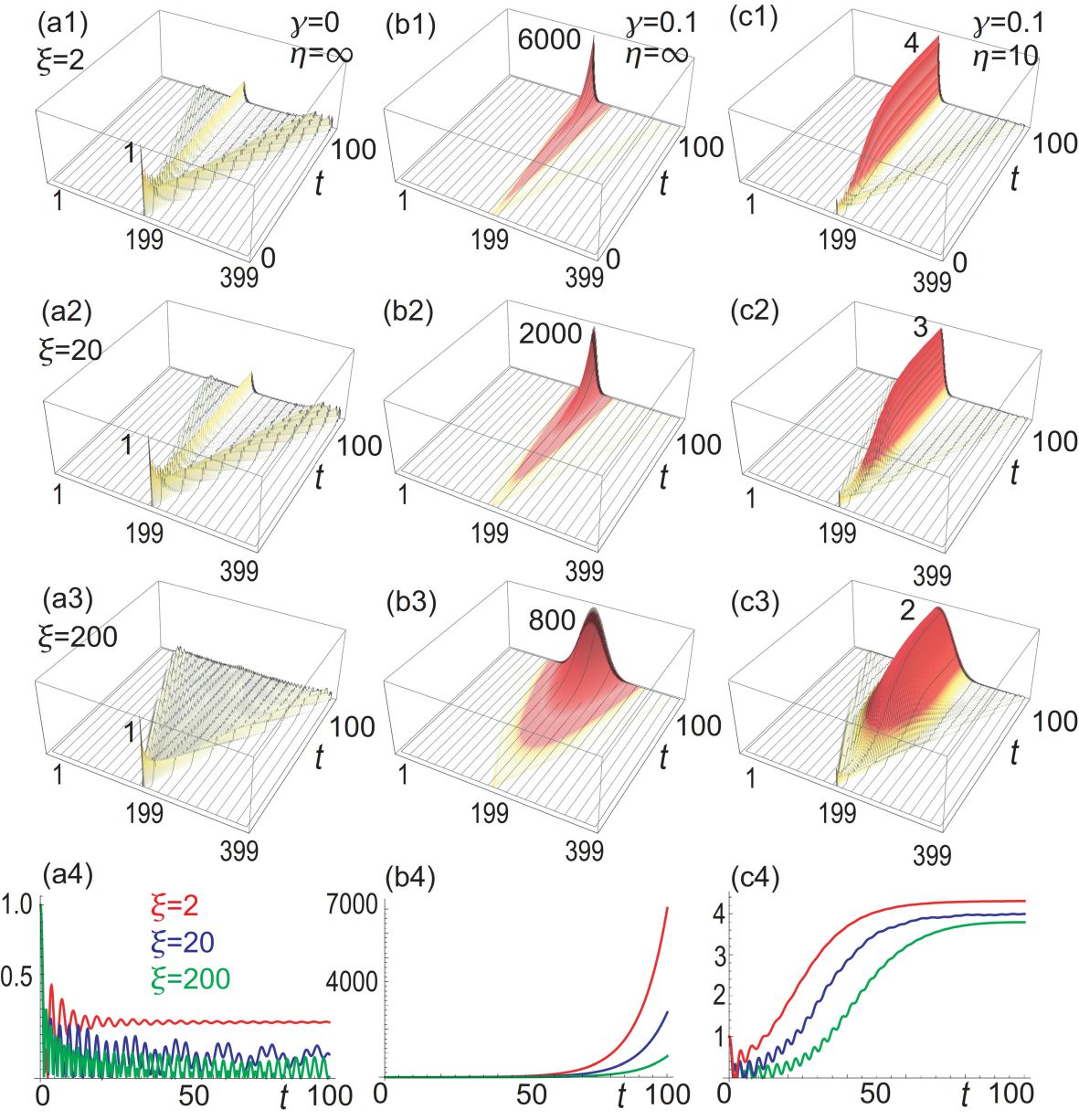}}
\caption{(*1), (*2), (*3) Time evolution of the spatial profile for the time
interval $0<t<100$ and (*4) that of the amplitude $\left\vert \protect\psi _{%
\text{IF}}\right\vert $\ at the interface for the time interval $0<t<100$
for various penetration depth ($\protect\xi =2$, $\protect\xi =20$, $\protect%
\xi =200$). (a*) Hermitian model ($\protect\gamma =0$). (b*) Linear
non-Hermitian model ($\protect\gamma =0.1$, $\protect\chi =2$, $\protect\eta %
=\infty $). (c*) Nonlinear non-Hermitian model ($\protect\gamma =0.1$, $%
\protect\chi =2$, $\protect\eta =10$). We have set $N=399$, where $n_{\text{%
IF}}=199$.}
\label{FigLinearWave}
\end{figure}

We now investigate the quench dynamics of the system by imposing the initial
condition (\ref{IniCon}).

First, we neglect the loss and gain terms by setting $\gamma =0$. We
numerically solve a set of differential equations (\ref{SSH1}) and (\ref%
{SSH2}), whose results are shown in Fig.\ref{FigLinearWave}(a1), (a2) and
(a3). The input given initially at the site $n=n_{\text{IF}}$ spreads over
the chain, but the component $|c_{\text{IF}}|$\ remains as it is, because $%
\phi _{\text{IF}}\left( t\right) =\phi _{\text{IF}}$\ in Eq.(\ref{TimeEvoIF}%
) for $\gamma =0$. There is a peak at the interface for $\xi =2$ as in Fig.%
\ref{FigLinearWave}(a1) but the peak is tiny for $\xi =200$ as in Fig.\ref%
{FigLinearWave}(a3).

Second, we include the linear loss and gain terms ($\gamma \chi \neq 0$),
whose results are shown in Fig.\ref{FigLinearWave}(b1), (b2) and (b3). The
topological interface state has a maximum value at the site with gain. As a
result, the state exponentially evolves and becomes infinite. However, this
is not physical. Indeed, there is a saturation of the gain in actual
experiments, about which we discuss in Section \ref{SecGain}. It is a
nonlinear saturation effect ($\eta <\infty $). Here, we present the results
in Fig.\ref{FigLinearWave}(c1), (c2) and (c3)\ by choosing $\eta =10.$

We show the time evolution of the amplitude $|\psi _{n_{\text{IF}}}|$ in Fig.%
\ref{FigLinearWave}(a4), (b4) and (c4). It becomes stationary after a
certain time in the absence of the loss and gain terms ($\gamma =0$) as
shown in Fig.\ref{FigLinearWave}(a4). On the other hand, the amplitude
exponentially becomes large once the loss and gain terms are present ($%
\gamma \chi \neq 0$), as shown in Fig.\ref{FigLinearWave}(b4). It becomes
stationary by the saturation term ($\eta <\infty $) as in Fig.\ref%
{FigLinearWave}(c4), about which we discuss in Section \ref{SecGain}.

\section{Jackiw-Rebbi solution in non-Hermitian model}

\label{SecDirac}

Supersymmetric quantum mechanics is a method to obtain an analytic solution
originally proposed by Witten\cite{Witten,CooperP,Cooper,Junker}. It has
also been applied to laser systems\cite%
{Heinrich,Miri,Gana,SUSYLaser,Mid,Qiao}.

We continue to study the linear model but based on the PT-symmetric
non-Hermitian SSH model $\overline{M}_{nm}$ from now. The two matrices $%
\widetilde{M}_{nm}$ and $\overline{M}_{nm}$\ are different only by a
c-number as in Eq.(\ref{HoppiMatrixB}). Hence, the eigenfunctions are
identical with the eigenvalues different only by this c-number.

We diagonalize the matrix $\overline{M}_{nm}$ by employing an approximation
similar to the one made by Jackiw and Rebbi. The hopping amplitude (\ref{EqA}%
) becomes constant as in Eq.(\ref{EqB}) far away from the interface. Then,
the hopping matrix $\overline{M}_{nm}$ can be presented in the momentum
space as%
\begin{equation}
\overline{H}\equiv \left( 
\begin{array}{cc}
i\bar{\gamma} & \kappa _{A}+\kappa _{B}e^{-iak} \\ 
\kappa _{A}+\kappa _{B}e^{iak} & -i\bar{\gamma}%
\end{array}%
\right) .
\end{equation}%
The energy spectrum reads%
\begin{equation}
\overline{E}\left( k\right) =\pm \sqrt{\kappa _{A}^{2}+\kappa
_{B}^{2}+2\kappa _{A}\kappa _{B}\cos ak-\bar{\gamma}^{2}},
\end{equation}%
which has a Dirac-like dispersion in the vicinity of the momentum $k=\pi /a$%
. Assuming a sufficiently smooth configuration in the vicinity of $k=\pi /a$%
, we expand it as%
\begin{equation}
\overline{H}=\left( 
\begin{array}{cc}
i\bar{\gamma} & \Delta _{0}+i\kappa k^{\prime } \\ 
\Delta _{0}-i\kappa k^{\prime } & -i\bar{\gamma}%
\end{array}%
\right) ,  \label{HDirac}
\end{equation}%
with%
\begin{equation}
\Delta _{0}=\kappa _{A}-\kappa _{B},\quad k^{\prime }=k-\pi .  \label{kPi}
\end{equation}%
We bring back this Hamiltonian to the continuous coordinate space as%
\begin{equation}
\overline{H}=\left( 
\begin{array}{cc}
i\bar{\gamma} & \Delta \left( x\right) -\kappa \partial _{x} \\ 
\Delta \left( x\right) +\kappa \partial _{x} & -i\bar{\gamma}%
\end{array}%
\right) =\left( 
\begin{array}{cc}
i\bar{\gamma} & A^{\dagger } \\ 
A & -i\bar{\gamma}%
\end{array}%
\right) ,  \label{HDiracRe}
\end{equation}%
with%
\begin{equation}
A\equiv \Delta \left( x\right) +\kappa \partial _{x},\qquad A^{\dagger
}\equiv \Delta \left( x\right) -\kappa \partial _{x},  \label{OpeA}
\end{equation}%
and%
\begin{equation}
\Delta \left( x\right) =\kappa \lambda \tanh \frac{x-x_{\text{IF}}}{a\xi },
\label{DeltaX}
\end{equation}%
where we have recovered the site dependent hopping amplitude from Eq.(\ref%
{EqA}).

The eigenequation of the Hamiltonian for $p$th eigenindex (\ref{HDiracRe})
reads%
\begin{equation}
\overline{H}\left( 
\begin{array}{c}
\Psi _{p}^{A}\left( x\right) \\ 
\Psi _{p}^{B}\left( x\right)%
\end{array}%
\right) =\overline{E}_{p}\left( 
\begin{array}{c}
\Psi _{p}^{A}\left( x\right) \\ 
\Psi _{p}^{B}\left( x\right)%
\end{array}%
\right) ,  \label{AGamma}
\end{equation}%
with (\ref{HDiracRe}), where we have defined the wavefunctions with the
eigenvalue $\overline{E}_{p}$ at the A and B sites as $\Psi ^{A}\left(
x\right) $ and $\Psi ^{B}\left( x\right) $, respectively.

\begin{figure}[t]
\centerline{\includegraphics[width=0.48\textwidth]{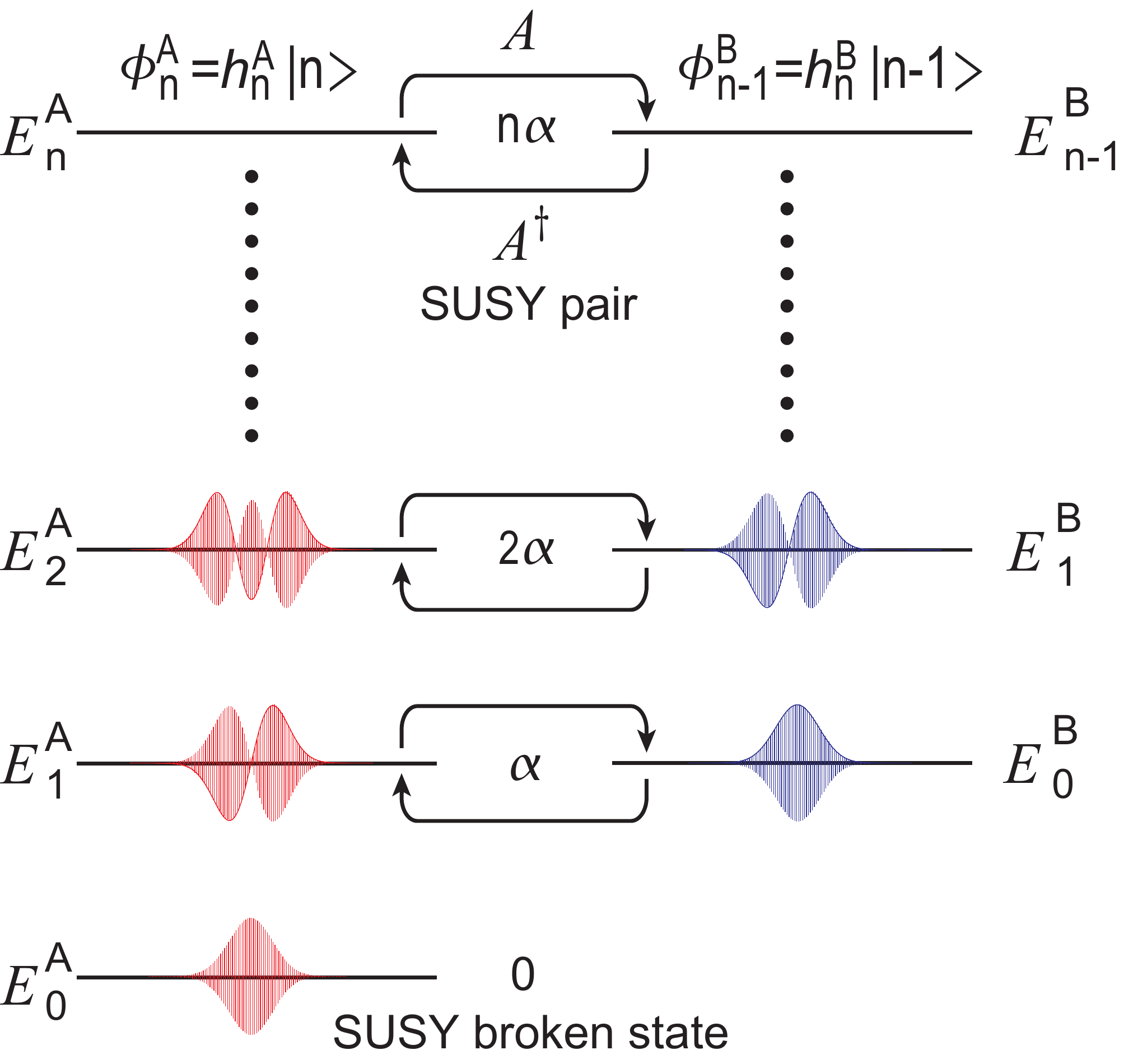}}
\caption{Illustration of the energy levels and the SUSY quantum mechanics.
Wave functions are shown in the case of $h_{n}^{A}=h_{n}^{B}$ for the
Hermitian system. }
\label{FigSUSYLevel}
\end{figure}

We derive the eigenfunction representing the topological interface state.
Its eigenenergy $\widetilde{E}_{\text{IF}}$ is given by Eq.(\ref{EneIF}) in
the $\widetilde{M}_{nm}$ basis, which reads $\overline{E}_{\text{IF}}=i\bar{%
\gamma}$ in the $\overline{H}$ basis. Hence, Eq.(\ref{AGamma}) yields%
\begin{equation}
\overline{H}\left( 
\begin{array}{c}
\Psi _{0}^{A}\left( x\right) \\ 
\Psi _{0}^{B}\left( x\right)%
\end{array}%
\right) =i\bar{\gamma}\left( 
\begin{array}{c}
\Psi _{0}^{A}\left( x\right) \\ 
\Psi _{0}^{B}\left( x\right)%
\end{array}%
\right) ,
\end{equation}%
with $\overline{E}_{0}=\overline{E}_{\text{IF}}=i\bar{\gamma}$ and (\ref%
{HDiracRe}) for $\overline{H}$. It is easy to obtain one solution by setting 
$\Psi _{0}^{B}\left( x\right) =0$. The equation for $\Psi ^{A}\left(
x\right) $ reads%
\begin{equation}
A\Psi _{0}^{A}\left( x\right) =\left[ \Delta \left( x\right) +\kappa
\partial _{x}\right] \Psi _{0}^{A}\left( x\right) =0,  \label{EqD}
\end{equation}%
for which the Jackiw-Rebbi solution follows, 
\begin{align}
\Psi _{0}^{A}\left( x\right) & =c\exp \left[ -\frac{1}{\kappa }%
\int^{x}\Delta \left( x^{\prime }\right) dx^{\prime }\right] ,  \label{JRa}
\\
\Psi _{0}^{B}\left( x\right) & =0,  \label{JRb}
\end{align}%
with $c$\ is a normalization constant. This is a non-Hermitian
generalization of the Jackiw-Rebbi mode with a pure imaginary eigenvalue. It
is the unique solution because there is no degeneracy in the topological
interface state.

\begin{figure}[t]
\centerline{\includegraphics[width=0.44\textwidth]{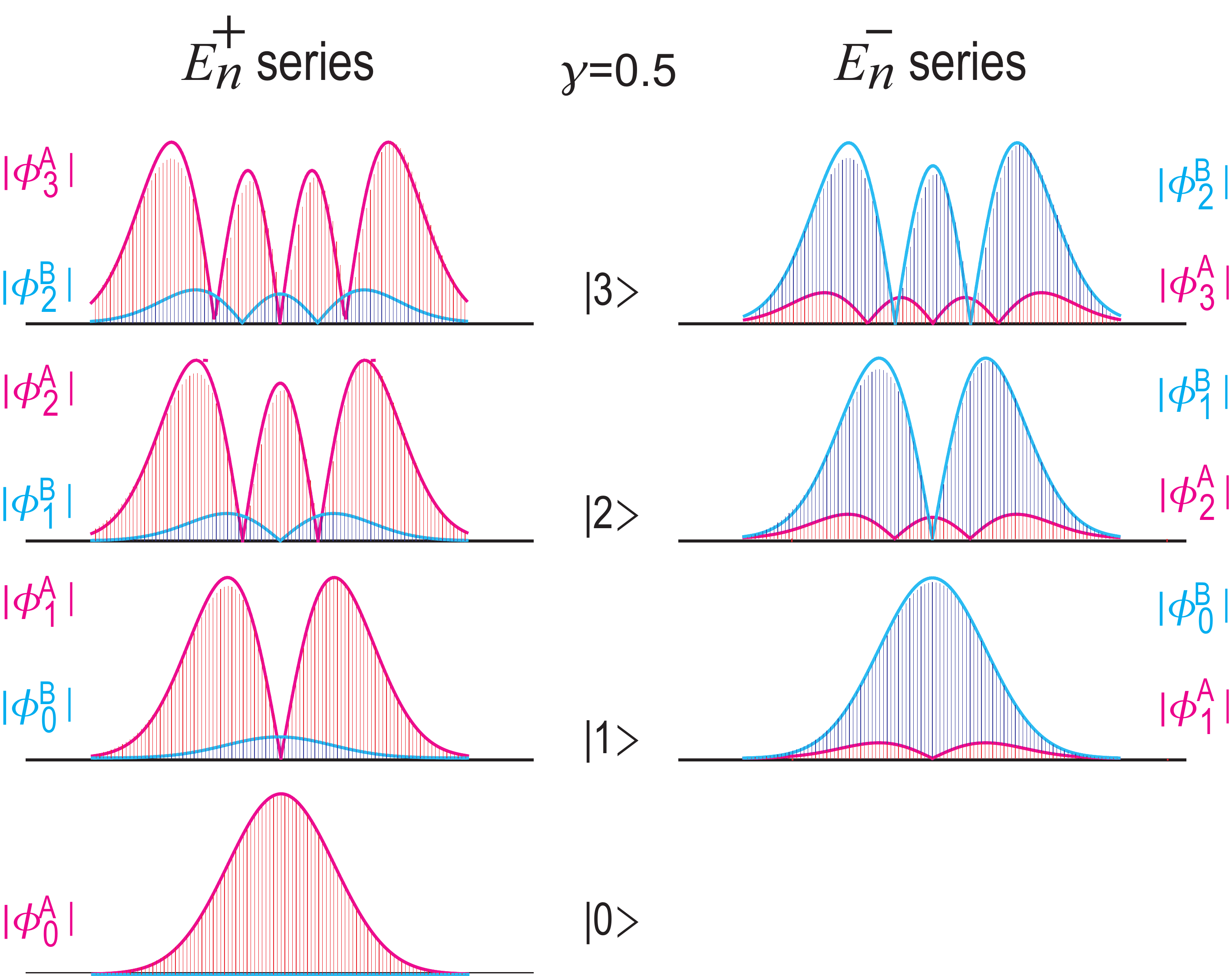}}
\caption{Red (blue) bars show the amplitudes numerically calculated at the A
(B) site. Magenta (cyan) heavy curves are analytical results given by Eq.(%
\protect\ref{HermPoly}), which envelop the numerical results very well. Each
panel contains SUSY partners made of amplitudes $|\protect\phi _{n}^{A}| $
and $|\protect\phi _{n-1}^{B}|$. Their magnitudes are quite different for $%
\protect\gamma =0.5$. The left (right) column is for the series of the
energy $\overline{E}_{n}^{+}$ ($\overline{E}_{n}^{-}$). }
\label{FigNHSUSYWave}
\end{figure}

\section{SUSY quantum mechanics}

When an operator $A$ is given, we may define the supercharges $Q$, $%
Q^{\dagger }$ and the Hamiltonian $\hat{H}$ by\cite%
{Witten,CooperP,Cooper,Junker}%
\begin{align}
Q &\equiv \left( 
\begin{array}{cc}
0 & 0 \\ 
A & 0%
\end{array}%
\right) ,\qquad Q^{\dagger }\equiv \left( 
\begin{array}{cc}
0 & A^{\dagger } \\ 
0 & 0%
\end{array}%
\right) , \\
\hat{H} &=\left\{ Q,Q^{\dagger }\right\} =\left( 
\begin{array}{cc}
A^{\dagger }A & 0 \\ 
0 & AA^{\dagger }%
\end{array}%
\right) .
\end{align}%
The superalgebra follows.%
\begin{equation}
\left\{ Q,Q\right\} =\left\{ Q^{\dagger },Q^{\dagger }\right\} =\left[ \hat{H%
},Q\right] =\left[ \hat{H},Q^{\dagger }\right] =0.
\end{equation}%
A representation of the algebra is constructed as follows.

We define the operators%
\begin{equation}
H_{A}\equiv A^{\dagger }A,\qquad H_{B}\equiv AA^{\dagger }.
\end{equation}%
The eigenvalue equations are 
\begin{equation}
H_{A}\phi _{p}^{A}=E_{p}^{A}\phi _{p}^{A},\qquad H_{B}\phi
_{p}^{B}=E_{p}^{B}\phi _{p}^{B}.  \label{EqAA}
\end{equation}%
Using these we obtain%
\begin{align}
H_{B}(A\phi _{p}^{A})& =AA^{\dagger }A\phi _{p}^{A}=E_{p}^{A}(A\phi
_{p}^{A}), \\
H_{A}(A^{\dagger }\phi _{p}^{B})& =A^{\dagger }AA^{\dagger }\phi
_{p}^{B}=E_{p}^{B}(A^{\dagger }\phi _{p}^{B}),
\end{align}%
and hence, $A\phi _{q}^{A}$ is an eigenstate of $H_{B}$ with the eigenvalue $%
E_{q}^{A}$. If we assume $E_{0}^{A}=0$ and $E_{0}^{B}\neq 0$, we may choose $%
q=p+1$. Then, $\phi _{p}^{B}\left( x\right) \varpropto A\phi
_{p+1}^{A}\left( x\right) $ and $\phi _{p+1}^{A}\left( x\right) \varpropto
A^{\dagger }\phi _{p}^{B}\left( x\right) $ so that%
\begin{equation}
E_{p}^{B}=E_{p+1}^{A},\qquad E_{0}^{A}=0.
\end{equation}%
The wavefunctions give a representation of the SUSY algebra, as illustrated
in Fig.\ref{FigSUSYLevel}.

We now show that the present model presents a non-Hermitian representation
of the SUSY algebra. We may use Eq.(\ref{OpeA}) for $A$ and $A^{\dagger }$.
For the Hamiltonian (\ref{HDiracRe}) we find%
\begin{equation}
\overline{H}^{2}=\left( 
\begin{array}{cc}
-\bar{\gamma}^{2}+A^{\dagger }A & 0 \\ 
0 & -\bar{\gamma}^{2}+AA^{\dagger }%
\end{array}%
\right) .
\end{equation}%
On the other hand, from%
\begin{equation}
\overline{H}^{2}\left( 
\begin{array}{c}
\Psi _{p}^{A}\left( x\right) \\ 
\Psi _{p}^{B}\left( x\right)%
\end{array}%
\right) =\overline{E}_{n}^{2}\left( 
\begin{array}{c}
\Psi _{p}^{A}\left( x\right) \\ 
\Psi _{p}^{B}\left( x\right)%
\end{array}%
\right) ,
\end{equation}%
we find a set of decoupled equations, 
\begin{align}
H_{A}\Psi _{p}^{A}\left( x\right) & =(\overline{E}_{p}^{2}+\bar{\gamma}%
^{2})\Psi _{p}^{A}\left( x\right) ,  \label{EqC} \\
H_{B}\Psi _{p}^{B}\left( x\right) & =(\overline{E}_{p}^{2}+\bar{\gamma}%
^{2})\Psi _{p}^{B}\left( x\right) .  \label{EqCC}
\end{align}%
When we set%
\begin{align}
\phi _{p}^{A}& =\Psi _{p}^{A}\left( x\right) ,\qquad \phi _{p-1}^{B}=\Psi
_{p}^{B}\left( x\right) ,  \label{EqH} \\
E_{p-1}^{B}& =E_{p}^{A}=\overline{E}_{p}^{2}+\bar{\gamma}^{2},
\end{align}%
Eq.(\ref{EqAA}) is satisfied. Hence, the SUSY algebra is satisfied. The SUSY
partners are the wavefunctions on the A and B sites.

\section{Explicit solutions of non-Hermitian SSH model}

We next seek the explicit solutions of the non-Hermitian model (\ref{AGamma}%
). This can be done by simplifying the function (\ref{DeltaX}). When $\xi $\
is large, we can approximate the gap function by a linear function as%
\begin{equation}
\Delta \left( x\right) =(\kappa \lambda /\xi )x,
\end{equation}%
where we set $x=0$ at the interface. The Hamiltonian is given by%
\begin{equation}
\overline{H}=\left( 
\begin{array}{cc}
i\bar{\gamma} & A^{\dagger } \\ 
A & -i\bar{\gamma}%
\end{array}%
\right) ,  \label{Hamil}
\end{equation}%
or%
\begin{equation}
\widetilde{H}=\left( 
\begin{array}{cc}
i\gamma \chi -i\gamma & A^{\dagger } \\ 
A & -i\gamma%
\end{array}%
\right) ,  \label{Htil}
\end{equation}%
where%
\begin{equation}
A\equiv (\kappa \lambda /\xi )x+\kappa \partial _{x},\qquad A^{\dagger
}\equiv (\kappa \lambda /\xi )x-\kappa \partial _{x}.
\end{equation}%
The commutator of the SUSY operators is calculated as%
\begin{equation}
\left[ A,A^{\dagger }\right] =\alpha ,\quad \text{with}\quad \alpha \equiv
2\kappa ^{2}\lambda /\xi .
\end{equation}%
The standard commutation relation of the annihilation and creation operators
follows,%
\begin{equation}
\left[ b,b^{\dagger }\right] =1,  \label{CCR}
\end{equation}%
in terms of the scaled operators $b$ and $b^{\dagger }$ defined by%
\begin{equation}
A\equiv \sqrt{\alpha }b,\quad A^{\dagger }\equiv \sqrt{\alpha }b^{\dagger }.
\end{equation}%
Eqs.(\ref{EqC}) and (\ref{EqCC}) are rewritten as%
\begin{equation}
\alpha b^{\dagger }b\Psi _{p}^{A}=(\overline{E}_{p}^{2}+\bar{\gamma}%
^{2})\Psi _{p}^{A},\quad \alpha (1+b^{\dagger }b)\Psi _{p}^{B}=(\overline{E}%
_{p}^{2}+\bar{\gamma}^{2})\Psi _{p}^{B}.
\end{equation}%
These are solved as%
\begin{align}
\Psi _{p}^{A}\left( x\right) & =h_{p}^{A}\langle x|p\rangle ,\quad \Psi
_{p}^{B}\left( x\right) =h_{p}^{B}\langle x|p-1\rangle ,  \label{EqI} \\
\overline{E}_{p}^{2}& =-\bar{\gamma}^{2}+\alpha p,  \label{EqII}
\end{align}%
for $p\geq 1$, where $h_{p}^{A}$ and $h_{p}^{B}$ are c-numbers. For $p=0$,
we have $\overline{E}_{0}=i\bar{\gamma}$, and the wavefunctions are given by
the non-Hermitian Jackiw-Rebbi solutions (\ref{JRa}) and (\ref{JRb}).

We note that the energy $\overline{E}_{p}$ of the $p$th level is pure
imaginary when%
\begin{equation}
p<\frac{\bar{\gamma}^{2}}{\alpha }=\frac{\gamma \chi \xi }{4\kappa
^{2}\lambda }.
\end{equation}%
We call the mode $|p\rangle $ the SUSY Jackiw-Rebbi mode, because we create
it from the Jackiw-Rebbi mode $|0\rangle $ by the operation of $b^{\dagger }$%
. The SUSY Jackiw-Rebbi modes are supersymmetric, while the Jackiw-Rebbi
mode breaks it.

\begin{figure*}[t]
\centerline{\includegraphics[width=0.88\textwidth]{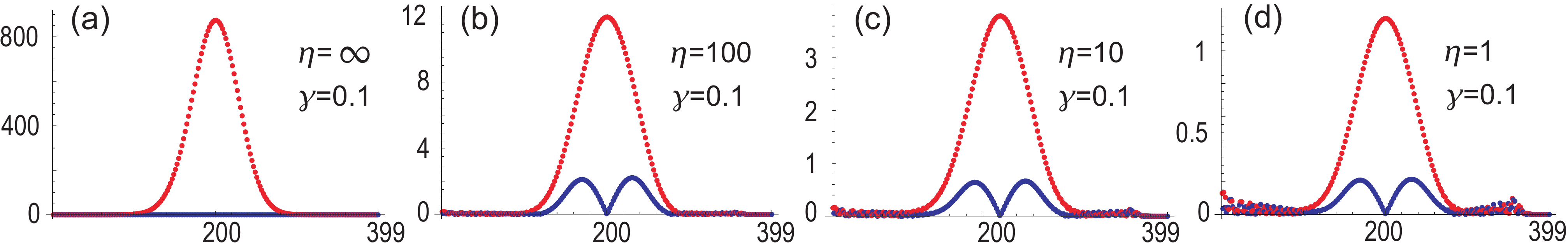}}
\caption{Spatial profile of the amplitude $|\Psi _{n}|$. (a) linear model,
(b) $\protect\eta =100$, (c) $\protect\eta =10$ and (d) $\protect\eta =1$.
We have set $\protect\chi =2$, $\protect\gamma =0.1$, $L=399$, $\protect\xi %
=200$ and $t=100$. The wavefunction is saturated and fixed real at the A
sites in red and pure imaginary at the B sites in blue for finite $\protect%
\eta $. }
\label{FigSaturation}
\end{figure*}

We determine the relation between two c-numbers $h_{p}^{A}$ and $h_{p}^{B}$.
We write down the eigenvalue equations (\ref{AGamma}) explicitly,%
\begin{equation}
i\bar{\gamma}\Psi _{p}^{A}+A^{\dagger }\Psi _{p}^{B}=\overline{E}_{p}\Psi
_{p}^{A},\quad A\Psi _{p}^{A}-i\bar{\gamma}\Psi _{p}^{B}=\overline{E}%
_{p}\Psi _{p}^{B},
\end{equation}%
which we rewrite with the use of (\ref{EqI}) as%
\begin{align}
i\bar{\gamma}h_{p}^{A}|p\rangle +\sqrt{\alpha }h_{p}^{B}b^{\dagger
}|p-1\rangle & =\overline{E}_{p}h_{p}^{A}|p\rangle , \\
\sqrt{\alpha }h_{p}^{A}b|p\rangle -i\bar{\gamma}h_{p}^{B}|p-1\rangle & =%
\overline{E}_{p}h_{p}^{B}|p-1\rangle .
\end{align}%
It follows that%
\begin{align}
i\bar{\gamma}h_{p}^{A}+\sqrt{\alpha p}h_{p}^{B}& =\overline{E}_{p}h_{p}^{A},
\\
\sqrt{\alpha p}h_{p}^{A}-i\bar{\gamma}h_{p}^{B}& =\overline{E}_{p}h_{p}^{B}.
\end{align}%
or%
\begin{align}
\sqrt{\alpha p}h_{p}^{B}& =(\overline{E}_{p}-i\bar{\gamma})h_{p}^{A}, \\
\sqrt{\alpha p}h_{p}^{A}& =(\overline{E}_{p}+i\bar{\gamma})h_{p}^{B},
\end{align}%
which leads to%
\begin{equation}
h_{p}^{B}=\left( \frac{\overline{E}_{p}-i\bar{\gamma}}{\overline{E}_{p}+i%
\bar{\gamma}}\right) ^{1/2}h_{p}^{A}.  \label{hBA}
\end{equation}%
Hence, the wavefunction $\Psi _{p}^{B}$ is determined once the wavefunction $%
\Psi _{p}^{A}$ is given.

Here we recall that there are two series of eigenfunctions corresponding to $%
\overline{E}_{p}^{\pm }=\pm \sqrt{-\bar{\gamma}^{2}+\alpha p}$ for $p\geq 1$
and $\overline{E}_{0}^{+}=i\bar{\gamma}$. We focus on SUSY Jackiw-Rebbi
modes, where $\bar{\gamma}^{2}>\alpha p$. In the parameter region with $\bar{%
\gamma}^{2}\gg \alpha p$, we would expand $\overline{E}_{p}^{\pm }=\pm i\bar{%
\gamma}+\cdots $. Then, we have%
\begin{align}
\left( \frac{h_{p}^{B}}{h_{p}^{A}}\right) ^{2}& =\frac{\overline{E}_{p}^{+}-i%
\bar{\gamma}}{\overline{E}_{p}^{+}+i\bar{\gamma}}\ll 1\quad \text{for}\quad 
\overline{E}_{p}^{+}=i\bar{\gamma}+\cdots , \\
\left( \frac{h_{p}^{B}}{h_{p}^{A}}\right) ^{2}& =\frac{\overline{E}_{p}^{-}-i%
\bar{\gamma}}{\overline{E}_{p}^{-}+i\bar{\gamma}}\gg 1\quad \text{for}\quad 
\overline{E}_{p}^{-}=-i\bar{\gamma}+\cdots .
\end{align}%
Hence,%
\begin{align}
|\Psi _{p}^{A}|& \gg |\Psi _{p}^{B}|\quad \text{for the series}\quad 
\overline{E}_{p}^{+}, \\
|\Psi _{p}^{A}|& \ll |\Psi _{p}^{B}|\quad \text{for the series}\quad 
\overline{E}_{p}^{-}.  \label{AmpRel}
\end{align}%
This explains a huge difference numerically found between\ the amplitudes at
the A and B sites in Fig.\ref{FigNHSUSYWave}.

We comment on the SUSY quantum mechanics. First of all, there are two series
of energies $\overline{E}_{p}^{\pm }=\pm \sqrt{-\bar{\gamma}^{2}+\alpha p}$,
although the relevant energies are $E_{p-1}^{B}=E_{p}^{A}=\alpha p$ for both
the series in SUSY quantum mechanics. However, the magnitudes of the
amplitudes are very different,%
\begin{align}
|\phi _{p}^{A}|& \gg |\phi _{p-1}^{B}|\quad \text{for the series}\quad 
\overline{E}_{p}^{+}, \\
|\phi _{p}^{A}|& \ll |\phi _{p-1}^{B}|\quad \text{for the series}\quad 
\overline{E}_{p}^{-},
\end{align}%
which follows from (\ref{EqH}) and (\ref{AmpRel}). These two series are
shown in Fig.\ref{FigNHSUSYWave}.

The wavefunction is given by $\langle x|p\rangle $ apart from the
normalization constant, and hence it is written in terms of the Hermite
polynomials precisely as in the Hermitian model,%
\begin{align}
\Psi _{p}^{A}\left( x\right) & =h_{p}^{A}\sqrt{\frac{1}{p!2^{p}}\sqrt{\frac{%
\lambda }{\pi \xi }}}H_{p}\left( \sqrt{\frac{\lambda }{\xi }}x\right) \exp %
\left[ -\frac{\lambda }{2\xi }x^{2}\right] , \\
\Psi _{p}^{B}\left( x\right) & =h_{p}^{B}\Psi _{p-1}^{A}\left( x\right) ,
\label{HermPoly}
\end{align}%
where $h_{p}^{B}$ is given by Eq.(\ref{hBA}) while $h_{p}^{A}$ is to be
determined numerically.

There is the Jackiw-Rebbi mode only for A site, whose wavefunctions are%
\begin{equation}
\Psi _{0}^{A}\left( x\right) =h_{0}^{A}\exp \left[ -\frac{\lambda }{2\xi }%
x^{2}\right] ,\quad \Psi _{0}^{B}\left( x\right) =0.  \label{SoluJR}
\end{equation}%
This is the SUSY-broken state.

Finally, we compare the analytic solutions and the numerical solutions in
Fig.\ref{FigNHSUSYWave}. The coincidence is very well between the analytic
solution and the numerical results except for a minor difference, where the
mirror symmetry is slightly broken in the numerical results. It is due to
the difference between the hopping parameters $\kappa _{A,n}$ and $\kappa
_{B}$ in Eq.(\ref{EqA}), where the band widths are different between the
topological and trivial phases. This difference is taken care of in the
numerical calculation but ignored in the analytical study.

\section{Gain with nonlinear saturation}

\label{SecGain}

\subsection{Quench dynamics}

We have so far studied the linear model containing loss and gain. The
amplitude increases infinitely as time passes. Actually, there must be a
saturation effect in gain, which makes the amplitude finite. We include the
saturation effect by keeping $\eta $ finite in Eq.(\ref{Master}). We show
the results in Fig.\ref{FigLinearWave}(c1), (c2) and (c3). The amplitudes
remain finite due to the saturation effect. It is a topological interface
laser stabilized by nonlinear and non-Hermicity effects. We also show the
time evolution of the amplitude $|\psi _{n_{\text{IF}}}|$ in Fig.\ref%
{FigLinearWave}(c4).

We show the spatial profile of the saturated amplitude $\left\vert \psi
_{n}\right\vert $ for various $\eta $ in Fig.\ref{FigSaturation}. Main
excitations are localized at the A sites in the vicinity of the interface,
whose wavefunction is real. However, there are also excitations at the B
sites in the vicinity of the interface as in Fig.\ref{FigSaturation}(b), (c)
and (d), whose wavefunction is pure imaginary. Hence, the relative phases
between the A and B sites are fixed to be $\pm i$\ and hence it will serve
as a large area coherent laser.

\subsection{Nonlinear Jackiw-Rebbi theory}

We have numerically revealed the excitations at the B sites in the presence
of the saturation term. We now show that they form the Jackiw-Rebbi mode
generalized to the nonlinear regime. Replacing the linear gain term with the
nonlinear gain term in Eq.(\ref{Htil}), we have%
\begin{equation}
\left( 
\begin{array}{cc}
\frac{i\gamma \chi }{1+\left\vert \Psi _{A}\left( x\right) \right\vert
^{2}/\eta }-i\gamma  & A^{\dagger } \\ 
A & -i\gamma 
\end{array}%
\right) \left( 
\begin{array}{c}
\Psi _{A}\left( x\right)  \\ 
\Psi _{B}\left( x\right) 
\end{array}%
\right) =E\left( 
\begin{array}{c}
\Psi _{A}\left( x\right)  \\ 
\Psi _{B}\left( x\right) 
\end{array}%
\right) .  \label{NLJR}
\end{equation}%
We analyze a small excitation at the B sites. Using a mean-field
approximation, we obtain $\Psi _{A}\left( x\right) $\ and $\Psi _{B}\left(
x\right) $\ as%
\begin{align}
\Psi _{A}\left( x\right) & =c\exp \left[ -\frac{\kappa \lambda }{2\xi }%
\left( 1+c_{2}\right) x^{2}\right] ,  \label{PsiA} \\
\Psi _{B}\left( x\right) & =-ic\frac{x}{\eta }\frac{c_{2}\kappa \lambda }{%
\xi }\exp \left[ -\frac{\kappa \lambda }{2\xi }\left( 1+c_{2}\right) x^{2}%
\right] ,  \label{PsiB}
\end{align}%
where $c$\ is a normalization constant, and%
\begin{equation}
c_{2}=\frac{\gamma ^{2}\chi ^{2}\xi }{\kappa ^{2}\lambda }\left[ \frac{1}{%
1+\left\vert \overline{\Psi }_{A}\right\vert ^{2}/\eta }-\frac{1}{%
1+\left\vert \Psi _{A}\left( 0\right) \right\vert ^{2}/\eta }\right] ,
\end{equation}%
with $\overline{\Psi }_{A}$\ the mean of $\Psi _{A}\left( x\right) $. See
Appendix B for detailed derivation. We note that 
\begin{equation}
\frac{\Psi _{B}\left( x\right) }{\Psi _{A}\left( x\right) }=-ix\frac{\gamma
^{2}\chi ^{2}}{\kappa }\left[ \frac{1}{1+\left\vert \overline{\Psi }%
_{A}\right\vert ^{2}/\eta }-\frac{1}{1+\left\vert \Psi _{A}\left( 0\right)
\right\vert ^{2}/\eta }\right] .  \label{PsiAB}
\end{equation}%
The relative phases between the A and B are fixed to be $\pm i$.
Furthermore, this formula well explains three key properties of the
wavefunctions revealed in Fig.\ref{FigSaturation}: (1) $\left\vert \Psi
_{A}\left( x\right) \right\vert $\ is proportional to $\sqrt{\eta }$; (2) $%
\Psi _{B}\left( x\right) /\Psi _{A}\left( x\right) $\ is independent of $\eta 
$; $\left\vert \Psi _{A}\left( x\right) \right\vert =0$\ at $x=0.$

\section{Conclusion and Discussion}

We have explored a SUSY structure in the SSH model with a topological
interface as a model of topological interface laser with gain and loss. By
extending a SUSY quantum mechanics to non-Hermitian systems, we have found a
series of analytic solutions which extend the original Jackiw-Rebbi
solution. They have pure imaginary energies and their wavefunctions are
given by those of a harmonic oscillator. We also derived an analytic form of
the Jackiw-Rebbi mode in nonlinear regime by using a mean-field
approximation.

We have applied quench dynamics to investigate a topological interface laser
with gain and loss. However, it may be hard to observe the time evolution in
actual optical experiments because the time scale is too short. The same
physics is executed by the coupled-wave-guide arrays along the $z$\ direction%
\cite{Longhi}, simply by replacing time $t$\ by coordinate $z$\ in the
equation of motion.

We have developed an analysis based on the basic equation (\ref{Master}). On
the other hand, it is well known that the dynamics of a laser is described
by the rate equations. It is actually possible to derive Eq.(\ref{Master})
from the rate equations in a certain limit provided the carrier population
is saturated. See details for Appendix C.

Large-area single-mode lasers are realized by suppressing the appearance of
higher order modes. In order to increase the output power of the laser, it
is necessary to enlarge the emitting area. However, it causes the multi-mode
and degrades the brightness at the same time in general. There are several
proposals on the single mode laser using photonic crystals have been
reported by using double-lattice photonic-crystal resonators\cite{Yoshida},
accidental Dirac-point\cite{Chua,Contra} and Kekul\'{e} modulation\cite{Gao,Yang}
in the photonic lattice mostly over the past few years. Our results give a
deeper understanding of a large area single mode laser from a topological
interface\cite{Ishida}.

M.E is supported by CREST, JST (Grants No. JPMJCR20T2). N. Ishida is
supported by the Grants-in-Aid for Scientific Research from MEXT KAKENHI
(Grants. No. JP21J40088).\ Y. Ota is supported by the Grants-in-Aid for
Scientific Research from MEXT KAKENHI (Grants. Nos. 22H01994 and 22H00298).
S. Iwamoto is supported by CREST, JST (Grants No. JPMJCR19T1) and the
Grants-in-Aid for Scientific Research from MEXT KAKENHI (Grants. Nos.
22H00298 and 22H01994).

\appendix

\section{Topological property of the non-Hermitian SSH model}

We consider a homogeneous system. The Hamiltonian in the momentum space
corresponding to the hopping matrix (\ref{HoppiMatrixA}) is%
\begin{align}
\widetilde{H}& =\left( 
\begin{array}{cc}
-i\gamma \left( 1-\chi \right) & \kappa _{A}+\kappa _{B}e^{-iak} \\ 
\kappa _{A}+\kappa _{B}e^{iak} & -i\gamma%
\end{array}%
\right) ,  \notag \\
& =-i\gamma \left( 1-\frac{\chi }{2}\right) I_{2}+\overline{H}_{\text{SSH}}
\label{SSH-k}
\end{align}%
with $a$ the lattice constant and%
\begin{equation}
\overline{H}\equiv \left( 
\begin{array}{cc}
i\gamma \chi /2 & \kappa _{A}+\kappa _{B}e^{-iak} \\ 
\kappa _{A}+\kappa _{B}e^{iak} & -i\gamma \chi /2%
\end{array}%
\right) .  \label{SSHbar}
\end{equation}%
The Hamiltonian $\overline{H}_{\text{SSH}}$ is non-Hermitian for $\gamma
\neq 0$. The relation between the eigenenergy of the Hamiltonians (\ref%
{SSH-k}) and (\ref{SSHbar}) is%
\begin{equation}
\widetilde{E}=-i\gamma \left( 1-\frac{\chi }{2}\right) +\overline{E}_{\text{%
SSH}}.
\end{equation}%
The energy spectrum reads%
\begin{equation}
\overline{E}\left( k\right) =\pm \sqrt{\kappa _{A}^{2}+\kappa
_{B}^{2}+2\kappa _{A}\kappa _{B}\cos ak-\gamma ^{2}}.  \label{Ene}
\end{equation}%
Especially, we have%
\begin{equation}
\overline{E}\left( \pi /a\right) =\pm \sqrt{\left( \kappa _{A}-\kappa
_{B}\right) ^{2}-\gamma ^{2}}.
\end{equation}%
The system is the PT preserved phase for $\gamma <\left\vert \kappa
_{A}-\kappa _{B}\right\vert $, where the bulk energy is real even though the
system is non-Hermitian, while the system is the PT\ broken phase for $%
\gamma >\left\vert \kappa _{A}-\kappa _{B}\right\vert $, where the bulk
energy becomes pure imaginary for a certain range of the momentum $k$.

We recall that the PT symmetry operation is defined by 
\begin{equation}
PT=\sigma _{x}K,
\end{equation}%
with K the complex conjugation. Since we have%
\begin{equation}
PT\overline{H}\left( k\right) \left( PT\right) ^{-1}=\overline{H}\left(
k\right) ,
\end{equation}%
and hence $\overline{H}_{\text{SSH}}$ is a PT symmetric Hamiltonian.

The topological number is defined with respect to the Hamiltonian (\ref%
{SSHbar}). We define the right and left eigenvectors by%
\begin{equation}
H\left\vert \psi ^{\text{R}}\right\rangle =E\left\vert \psi ^{\text{R}%
}\right\rangle ,\quad H^{\dagger }\left\vert \psi ^{\text{L}}\right\rangle
=E\left\vert \psi ^{\text{L}}\right\rangle .
\end{equation}%
The non-Hermitian Zak phase is a topological number\cite{Lieu}%
\begin{equation}
W\equiv \frac{i}{2\pi /a}\int_{0}^{2\pi /a}\left\langle \psi ^{\text{L}%
}\right\vert \frac{\partial }{\partial k}\left\vert \psi ^{\text{R}%
}\right\rangle dk.
\end{equation}%
It is straightforward to show that $W=1$ for $\kappa _{A}<\kappa _{B}$ and $%
W=0$ for $\kappa _{A}>\kappa _{B}$ irrespective of $\gamma $. Hence, the
system is topological for $\kappa _{A}<\kappa _{B}$ and trivial for $\kappa
_{A}>\kappa _{B}$.

\section{Nonlinear Jackiw-Rebbi solution}

We derive a set of the saturated distribution (\ref{PsiA}) and (\ref{PsiB})
from Eq.(\ref{NLJR}). First, we write Eq.(\ref{NLJR}) explicitly as%
\begin{align}
& i\gamma \left( \frac{\chi }{1+\left\vert \Psi _{A}\left( x\right)
\right\vert ^{2}/\eta }-1\right) \Psi _{A}\left( x\right) +A^{\dagger }\Psi
_{B}\left( x\right) =E\Psi _{A}\left( x\right) , \\
& A\Psi _{A}\left( x\right) -i\gamma \Psi _{B}\left( x\right) =E\Psi
_{B}\left( x\right) ,
\end{align}%
where $A$\ and $A^{\dagger }$\ are given by Eq.(\ref{OpeA}) with Eq.(\ref%
{DeltaX}). The second equation is solved as%
\begin{equation}
\Psi _{B}\left( x\right) =\frac{A\Psi _{A}\left( x\right) }{E+i\gamma },
\label{EqJ}
\end{equation}%
which we insert into the first equation to derive 
\begin{align}
& A^{\dagger }A\Psi _{A}\left( x\right)  \notag \\
& =\left( E+i\gamma \right) \left[ E-i\gamma \left( \chi \frac{1}{%
1+\left\vert \Psi _{A}\left( x\right) \right\vert ^{2}/\eta }-1\right) %
\right] \Psi _{A}\left( x\right) .
\end{align}%
We assume that the energy is modified from Eq.(\ref{EneIFB}) as%
\begin{equation}
E=i\gamma \left( \chi -1\right) +c_{1},
\end{equation}%
where $c_{1}$\ is a constant to be determined. Inserting it and we have%
\begin{equation}
A^{\dagger }A\Psi _{A}\left( x\right) \simeq i\gamma \chi \left[
c_{1}+i\gamma \chi \left( 1-\frac{1}{1+\left\vert \Psi _{A}\left( 0\right)
\right\vert ^{2}/\eta }\right) \right] \Psi _{A}\left( x\right) ,  \notag
\end{equation}%
where we have used an approximation $\left\vert \Psi _{A}\left( x\right)
\right\vert ^{2}\simeq \left\vert \Psi _{A}\left( 0\right) \right\vert ^{2}$%
\ because $\Psi _{A}\left( x\right) $\ rapidly decreases except at $x=0$. We
choose%
\begin{equation}
c_{1}=i\gamma \chi \left( \frac{1}{1+\left\vert \overline{\Psi }%
_{A}\right\vert ^{2}/\eta }-1\right) ,
\end{equation}%
where $\overline{\Psi }_{A}$\ is the mean value of $\Psi _{A}\left( x\right) 
$. We obtain%
\begin{equation}
A^{\dagger }A\Psi _{A}\left( x\right) =-\gamma ^{2}\chi ^{2}\left[ \frac{1}{%
1+\left\vert \overline{\Psi }_{A}\right\vert ^{2}/\eta }-\frac{1}{%
1+\left\vert \Psi _{A}\left( 0\right) \right\vert ^{2}/\eta }\right] \Psi
_{A}\left( x\right) .  \label{AdAgg}
\end{equation}%
On the other hand, we assume a wavefunction modified from Eq.(\ref{SoluJR})
as%
\begin{equation}
\Psi _{A}\left( x\right) =c\exp \left[ -\frac{\kappa \lambda }{2\xi }\left(
1+c_{2}\right) x^{2}\right] ,
\end{equation}%
where $c$\ is a normalization constant and $c_{2}$\ is a constant to be
determined. Applying $A$\ and $A^{\dagger }A$\ to $\Psi _{A}\left( x\right) $%
, we obtain%
\begin{eqnarray}
A\Psi _{A}\left( x\right) &\simeq &-\frac{c_{2}\kappa \lambda }{\xi }x\Psi
_{A}\left( x\right) ,  \label{EqS} \\
A^{\dagger }A\Psi _{A}\left( x\right) &\simeq &\frac{c_{2}\kappa ^{2}\lambda 
}{\xi }\Psi _{A}\left( x\right) .  \label{EqT}
\end{eqnarray}%
Comparing (\ref{EqT}) with Eq.(\ref{AdAgg}), we obtain%
\begin{equation}
c_{2}=\frac{\gamma ^{2}\chi ^{2}\xi }{\kappa ^{2}\lambda }\left[ \frac{1}{%
1+\left\vert \overline{\Psi }_{A}\right\vert ^{2}/\eta }-\frac{1}{%
1+\left\vert \Psi _{A}\left( 0\right) \right\vert ^{2}/\eta }\right] .
\end{equation}%
With the use of Eqs.(\ref{EqJ}) and (\ref{EqS}), $\Psi _{B}\left( x\right) $%
\ is derived as%
\begin{equation}
\Psi _{B}\left( x\right) =-icx\frac{c_{2}\kappa \lambda }{\xi }\exp \left[ -%
\frac{\kappa \lambda }{2\xi }\left( 1+c_{2}\right) x^{2}\right] .
\end{equation}%
It is the saturated distribution (\ref{PsiB}) in the main text. We then have

\begin{equation}
\frac{\Psi _{B}\left( x\right) }{\Psi _{A}\left( x\right) }=-ix\frac{\gamma
^{2}\chi ^{2}}{\kappa }\left[ \frac{1}{1+\left\vert \overline{\Psi }%
_{A}\right\vert ^{2}/\eta }-\frac{1}{1+\left\vert \Psi _{A}\left( 0\right)
\right\vert ^{2}/\eta }\right] ,
\end{equation}%
which is Eq.(\ref{PsiAB}) in the main text.

\begin{widetext}

\section{Rate equation}

The rate equations read\cite{Hass,Parto}%
\begin{eqnarray}
\frac{dE_{n}^{A}}{dt} &=&\frac{1}{2}\left[ -\gamma _{0}+\sigma \left(
N_{n}^{A}-1\right) \right] \left( 1-i\alpha _{\text{H}}\right)
E_{n}^{A}+i\kappa _{A}^{0}E_{n}^{B}+i\kappa _{B}^{0}E_{n-1}^{B}, \\
\frac{dE_{n}^{B}}{dt} &=&\frac{1}{2}\left[ -\gamma _{0}+\sigma \left(
N_{n}^{B}-1\right) \right] \left( 1-i\alpha _{\text{H}}\right)
E_{n}^{B}+i\kappa _{A}^{0}E_{n}^{A}+i\kappa _{B}^{0}E_{n+1}^{A}, \\
\frac{dN_{n}^{A}}{dt} &=&R_{A}-\frac{N_{n}^{A}}{\tau _{r}}-F\left(
N_{n}^{A}-1\right) \left\vert E_{n}^{A}\right\vert ^{2}, \\
\frac{dN_{n}^{B}}{dt} &=&R_{B}-\frac{N_{n}^{B}}{\tau _{r}}-F\left(
N_{n}^{B}-1\right) \left\vert E_{n}^{B}\right\vert ^{2},
\end{eqnarray}%
where $E_{n}^{A}$ and $E_{n}^{B}$ are electric field amplitudes in
sublattices $A$ and $B$ and $N_{n}^{A}$ and $N_{n}^{B}$ are carrier
population densities.

We assume the carrier is saturated%
\begin{equation}
\frac{dN_{n}^{A}}{dt}=0,\qquad \frac{dN_{n}^{A}}{dt}=0,
\end{equation}%
or%
\begin{eqnarray}
N_{n}^{A}-1 &=&\frac{F\left\vert E_{n}^{A}\right\vert ^{2}+R_{A}}{%
F\left\vert E_{n}^{A}\right\vert ^{2}+1/\tau _{r}}-1=\frac{R_{A}-1/\tau _{r}%
}{F\left\vert E_{n}^{A}\right\vert ^{2}+1/\tau _{r}}, \\
N_{n}^{B}-1 &=&\frac{F\left\vert E_{n}^{B}\right\vert ^{2}+R_{B}}{%
F\left\vert E_{n}^{B}\right\vert ^{2}+1/\tau _{r}}-1=\frac{R_{B}-1/\tau _{r}%
}{F\left\vert E_{n}^{B}\right\vert ^{2}+1/\tau _{r}}.
\end{eqnarray}%
By inserting them into the rate equations, we have%
\begin{eqnarray}
\frac{dE_{n}^{A}}{dt} &=&\frac{1}{2}\left[ -\gamma _{0}+\sigma \frac{%
R_{A}-1/\tau _{r}}{F\left\vert E_{n}^{A}\right\vert ^{2}+1/\tau _{r}}\right]
\left( 1-i\alpha _{\text{H}}\right) E_{n}^{A}+i\kappa
_{A}^{0}E_{n}^{B}+i\kappa _{B}^{0}E_{n-1}^{B}, \\
\frac{dE_{n}^{B}}{dt} &=&\frac{1}{2}\left[ -\gamma _{0}+\sigma \frac{%
R_{B}-1/\tau _{r}}{F\left\vert E_{n}^{B}\right\vert ^{2}+1/\tau _{r}}\right]
\left( 1-i\alpha _{\text{H}}\right) E_{n}^{B}+i\kappa
_{A}^{0}E_{n}^{A}+i\kappa _{B}^{0}E_{n+1}^{A},
\end{eqnarray}%
or%
\begin{eqnarray}
i\frac{dE_{n}^{A}}{dt} &=&\frac{i}{2}\left[ -\gamma _{0}+\sigma \frac{\tau
_{r}R_{A}-1}{1+\tau _{r}F\left\vert E_{n}^{A}\right\vert ^{2}}\right] \left(
1-i\alpha _{\text{H}}\right) E_{n}^{A}-\kappa _{A}^{0}E_{n}^{B}-\kappa
_{B}^{0}E_{n-1}^{B}, \\
i\frac{dE_{n}^{B}}{dt} &=&\frac{i}{2}\left[ -\gamma _{0}+\sigma \frac{\tau
_{r}R_{B}-1}{1+\tau _{r}F\left\vert E_{n}^{B}\right\vert ^{2}}\right] \left(
1-i\alpha _{\text{H}}\right) E_{n}^{B}-\kappa _{A}^{0}E_{n}^{A}-\kappa
_{B}^{0}E_{n+1}^{A}.
\end{eqnarray}%
When $\alpha _{\text{H}}$ is negligible and $\tau _{r}R_{B}=1$, by setting%
\begin{eqnarray}
\psi _{n}^{A} &=&E_{n}^{A},\qquad \psi _{n}^{B}=E_{n}^{B},\qquad \kappa
_{A}=-\kappa _{A}^{0},\qquad \kappa _{B}=-\kappa _{B}^{0}, \\
\gamma &=&-\gamma _{0}/2,\qquad \eta =\tau _{r}F,\qquad \gamma \chi =\sigma
\left( \tau _{r}R_{A}-1\right) ,
\end{eqnarray}%
they are reduced to Eq.(\ref{Master}) in the main text.
\end{widetext}

\end{document}